\documentstyle[aps,12pt,manuscript]{revtex}
\begin{document}
\preprint{INJE--TP--96--5}
\def\overlay#1#2{\setbox0=\hbox{#1}\setbox1=\hbox to \wd0{\hss #2\hss}#1%
\hskip -2\wd0\copy1}

\title{ Higher dimensional extremal black strings}

\author{ H.W. Lee$^1$, Y.S. Myung$^1$, Jin Young Kim$^2$ and D.K. Park$^3$ }
\address{$^1$Department of Physics, Inje University, Kimhae 621--749, Korea\\
$^2$ Division of Basic Science, Dongseo University, Pusan 616--010, Korea \\
$^3$ Department of Physics, Kyungnam University, Masan 631--701, Korea} 

\maketitle

\vskip 1.5in

\begin{abstract}
We investigate  the five and six-dimensional black strings within Einstein-Maxwell
theory.  The extremal
black strings are endowed  with the null Killing symmetry.
We study the propagation of Einstein-Maxwell modes in the extremal black string
background  by using this symmetry. It turns out that one graviton is a propagating
mode, while both the Maxwell$({\cal F})$ and three-form $({\cal H})$ fields
are non-propagating modes. Further we discuss the stability  and classical hair
of the extremal black strings.
\end{abstract}

\newpage
\section{Introduction}
Recently there has been much progress in understanding the microscopic 
origin of
the black hole entropy. This was made possible by using a new description of 
solitonic states in 
string theory\cite{Vafa,Callan,Horo1,Breck,John,Horo2}.  
For the simplest five-dimensional extremal black hole, Strominger and 
Vafa \cite{Vafa} counted the number 
of degeneracy corresponding to BPS-saturated states in the string 
theory for given charge.
And they showed that 
for large charge, the number of states increases as $e^{A/4}$, where $A$ 
is the area of the horizon for extremal
black hole. That is, the statistical interpretation for the 
Bekenstein-Hawking entropy 
was made possible in five dimensions. Further, 
the five-dimensional black hole is just
a six-dimesional black string which winds around a compact internal 
circle\cite{Horo3}. 
The microstates of
five-dimensional extremal black hole are based on the fields moving 
around a circle in the internal
dimensions. In order to understand this situation, it is useful to take 
this internal direction as a space-time direction explicitly. This is a 
six dimensional black string solution. 
 
On the other hand, the four-dimensional extreme dilation black hole with
the coupling
constant $a= \sqrt{ p \over p+2}$ could be interpreted as a non-dilatonic, black
$p$-brane in $(4+p)$ dimensional Einstein-Maxwell system\cite{Gibb}. For example, for
$p=1$ one has the  $a=1/\sqrt 3$ dilaton black hole. Thus the four-dimensional extreme dilaton black hole
can be considered as the double-dimensional reduction of a non-singular
five-dimensional black string. The double dimensional reduction implies that
the spacetime dimension is reduced from five to four and simultaneously
 the the dimension of the extended object from one (black string) to zero
(black hole). The special case of $a=1/\sqrt 3$ is of particular interest because
the five-dimensional black string is a solution to the pure five-dimensional
supergravity. In this case one can derive a Bogomolnyi-type bound on the energy
 per unit length and this bound is saturated by the black string solution.
 We also note the relevance of this model to three-dimensional black string.

In this paper we study the higher dimensional black strings within Einstein-Maxwell
theory. We remind the reader that the  Einstein-Maxwell model
is the simplest one to 
study the higher dimensional black strings.
In order to relate these to the counting of the black hole entropy
one has to consider the ten-dimensional string theories\cite{Horo3}.
Actually the five-dimensional black string should be investigated in the context of
the type IIA string theory\cite{Horo4}. This theory  contains
the fields from the NS-NS sector: a metric ($G_{MN}$), a two-form
($B_{MN}$), and a dilaton ($\phi$)\cite{green}. Also the fields from the RR sector are
a one-form ($A_M$) and a three-form ($C_{MNP}$).
 This carries three charges: a magnetic charge 
 with respect to the RR two-form ($G=dA$), an electric charge with respect to 
the RR four-form ($F=dC$), and  a magnetic charge with respect to 
the NS-NS three-form ($H=dB$).
We consider the compactification from ten to four: $ M^{10} \to M^4 \times M^6$.
One takes six dimensions  to be  compactified to form a torus  and 
assumes translational symmetry in five of these dimensions $(M^6 \to 
S^1 \times \tilde S^1 \times \tilde T^4)$. The sixth direction along $S^1$
will have  a length$(L)$ much longer than  the others, and will be the direction
 in which the  waves propagate. Hence this solution corresponds to a black string
with traveling waves in five dimensions. Four toroidal directions ($\tilde T^4$)
form a torus of volume $\tilde {\cal V} = (2 \pi)^4 V$, and one along $\tilde S^1$ 
forms a circle of length $\tilde L$. The other four dimensions ($M^4$)
will be realized into a space-time.

For a complete study on the six-dimensional black string, one  has to consider
the ten-dimensional type IIB string theory compactified on a four torus 
 with volume ${\cal V}$. This ten-dimensional theory consists of (1) NS-NS sector :
a dilaton ($\varphi$), a metric ($g_{MN}$) and a two-form field ($B^{(1)}_{MN}$)
 (2) RR sector : an axion field ($\chi$), a two-form ($B^{(2)}_{MN}$), and
 a self dual four-form($\tilde C_{MNPQ}$).
This compactification  turns out to be ($M^{10} \to M^6 \times T^4$).
 We assume one additional spatial direction in six is toroidally compactified
 to form a circle ($S^1$) of length $L$ and choose $L >> {\cal V}^{ 1 \over 4}$
so that the solutions resemble strings in six dimensions. Here this solution
carries also   electric and magnetic charges with respect to the RR three-form
 ($H^{(2)}=dB^{(2)}$).  The object which carries a charge under the NS-NS
gauge field ($B^{(1)}_{MN}$) is  the elementary string itself.
At low energies the elementary string is described by a classical
solution of the effective field theory. However there are no states 
which carry charges of the RR gauge field ($B^{(2)}_{MN}$). 
Recently, it turns out that these missing states carrying  RR charges are D-branes\cite{Polch}.
We note that these  solutions with charges   are not precisely our case 
but  are very similar  to the kind of problems we are interested in.
As a first step to analyze the five and six-dimensional black strings generically,
we consider here the Einstein-Maxwell theory.

Garfinkle used a generating technique to find out the traveling waves of
the black string\cite{Garf1,Garf2}. Starting from a known static solution, this technique 
produces a new solution representing waves traveling on the old black string
background. 
However, this method has some limitations, since it always requires that 
the background metric ($\bar g_{MN}$)  possess a null,
orthogonal Killing vector.
 Further, Horowitz and Tseytlin  have worked on finding
traveling wave solutions of string theories in various dimensions\cite{Tsey1,Tsey2}.
They related a limit of their $F$-model to the extremal black string 
in three dimensions. Recently Horowitz and Marolf counted the number of microstates
for black string with traveling waves to obtain its entropy\cite{Horo4}.   
 
We use the standard scheme for studying the black hole as well as black string.
 Our method is  based on that for the black hole analysis\cite{Chan}.
But there are some differences.
 For the black 
holes, we choose the metric perturbation ($h_{\mu\nu}$) in such a way that 
the background symmetry should be restored at the perturbation 
level\cite{Kim}.  However, it is also necessary 
for the black strings\cite{Garf1,Garf2} to exploit such a background symmetry.
For the extremal black strings, the background symmetry corresponds to
the null Killing isometry. Furthermore in the black hole physics,
the counting of degrees of freedom for physical field is crucial for initial
setting of the perturbing fields. On the other hand,
it is more important for the extremal black string physics to use 
the null Killing vector field in choosing the perturbations. 

The  paper is organized as follows. In the next section, we will review the 
mathematical formalism for the subsequent  study. Especially we explain
the relationship between the dilaton black holes and higher dimensional 
black strings. In Sec.III we study the propagation of graviton and Maxwell
modes in five-dimensional black string background. The propagation of
six-dimensional black string is investigated in Sec.IV.  The last section offers 
conclusion and  discussion for our results. 

\section{Formalism}

Given a solution of the ($d+p$)-dimensional Einstein equations with a black 
$p$-brane (a horizon and $p$-fold translational symmetry), one can always find a
dilaton black hole solution of $d$-dimensional gravity by 
the double-dimensional reduction. The $(d+p)$-dimensional action is given by\cite{Gibb}
\begin{equation}
S_{d+p} = \int d^{(d+p)} x \sqrt{-g} 
  ( R_{d+p} - {2 \over (d-2)!} F^2_{d-2} ).
\end{equation}
We perform the double dimensional reduction of $p$ dimensions by taking
the $(d+p)$-dimensional metric and $(d-2)$-form $F^2_{d-2}$ to be
\begin{equation}
\tilde g_{MN} =
 \left(  \begin{array}{cc}  e^{2 \beta \phi(x)} g_{\mu \nu}  & 0  \\
              0 &  e^{2 \alpha \phi(x)} g_{mn}
\end{array}   \right)
\end{equation}
and
\begin{equation}
F_{(d-2)}= {1 \over (d-2)!} F_{\mu_1 \cdots \mu_{d-2}} dx^{\mu_1} \cdots 
          dx^{\mu_{d-2}},
\end{equation} 
where $x^\mu (y^m)$ are the coordinates for the $d$-dimensional spacetime
($p$-brane).
Note that both the metric and Maxwell field are independent of $y^m$. 
We do not include any $x$-$y$ cross term in the metric, which give rise to
additional gauge fields in $d$-dimensional spacetime. Using the relation
for the conformal transformation by $\tilde g_{MN} = \Omega^2 g_{MN}$,
we show that for $p \alpha + (d-2) \beta =0$,
(1) turns out to be the $d$-dimensional action.  After fixing the normalization
of the dilaton kinetic term as
\begin{equation}
\alpha^2 = { 2(d-2) \over p(d+p-2)},
\end {equation}
the $d$-dimensional dilaton action is given by
\begin{equation}
S_d = \int d^d x \sqrt{-g} 
  \Big \{ R_d - 2 (\nabla \phi)^2 - {2 \over (d-2)!}e^{- 2 a \phi} F^2_{d-2} 
     \Big \},
\end{equation}
where 
\begin{equation}
a = { (d-3) \sqrt{2p} \over \sqrt{(d-2)(d+p-2)}}.
\end {equation}
Then the magnetically charged black hole solutions to the Euler-Lagrange
 equations of (5) are

\begin{eqnarray}
&ds^2_d  &= - \Big [ 1 - \Big ( { r_+ \over r } \Big )^{d-3} \Big ]  
  \Big [ 1 - \Big ( { r_- \over r } \Big )^{d-3} 
                      \Big ]^{1 - (d-3) \gamma} dt^2  \nonumber   \\
&& + \Big [ 1 - \Big ( { r_+ \over r } \Big )^{d-3} \Big ]^{-1}  
  \Big [ 1 - \Big ( { r_- \over r } \Big )^{d-3} 
                      \Big ]^{ \gamma - 1} dr^2 
 + r^2 \Big [ 1 - \Big ( { r_- \over r } \Big )^{d-3} 
                      \Big ]^\gamma d \Omega^2_{(d-2)},     \\
& e^{a \phi} &= \Big [ 1 - \Big ( { r_- \over r } \Big )^{d-3} 
                      \Big ]^{ - { (d-3) \gamma \over 2} }, \nonumber   \\
& F_{(d-2)} &= Q \varepsilon_{d-2},   \nonumber
\end{eqnarray}
where $\varepsilon_{d-2}$ is the volume form for the unit $(d-2)$-sphere and
\begin{equation}
\gamma = { 2p \over (d-2)(p+1)}.
\end {equation}
Here a single charge $Q$ is related to the inner$(r_-)$ and outer $(r_+)$ 
horizons by
\begin{equation}
Q^2 = { (d+p-2)(d-3) \over 2(p+1)} (r_+r_-)^{d-3}.
\end {equation}
A black $p$-brane solution of the action (1) is given by

\begin{eqnarray}
&ds^2_{d + p}  &= - \Big [ 1 - \Big ( { r_+ \over r } \Big )^{d-3} \Big ]  
  \Big [ 1 - \Big ( { r_- \over r } \Big )^{d-3} 
                      \Big ]^{1 - p \over 1 + p} dt^2  
+\Big [ 1 - \Big ( { r_- \over r } \Big )^{d-3} 
                      \Big ]^{ 2 \over 1 + p} d {\bf y}^2_p
\nonumber   \\
&& + \Big [ 1 - \Big ( { r_+ \over r } \Big )^{d-3} \Big ]^{-1}  
  \Big [ 1 - \Big ( { r_- \over r } \Big )^{d-3} 
                      \Big ]^{ - 1} dr^2 
 + r^2 d \Omega^2_{(d-2)}.   
\end{eqnarray}
For the 
nonextremal case ($r_+>r_-$), both the black holes and black $p$-branes have
event horizons at $r=r_+$ and Cauchy horizons at $r=r_-$. 
However the situation is quite interesting for the 
extremal case  $(r_+=r_-= \mu)$.  In this case, the black hole
horizon becomes singular. On the other hand, the black $p$-brane solution leads to
\begin{equation}
ds^2_{d+p,ext}= \Big [ 1 - \Big ( { \mu \over r } \Big )^{d-3} 
      \Big ]^{ 2\over 1 + p} ( - dt^2 + d {\bf y}^2_p )
+ \Big [ 1 - \Big ( { \mu \over r } \Big )^{d-3} \Big ]^{-2} dr^2 
 + r^2 d \Omega^2_{(d-2)}.  
\end{equation}
This metric is invariant under the full $(p+1)$-Poincare group.
For appropriate values of $d$ and $p$, (11) includes all the known extreme, 
non-dilatonic, extented object solutions of higher dimensional supergravities.
These are membrane and fivebrane solutions of the  eleven-dimensional supergravity,
the self-dual three-brane of the ten-dimensional supergravity and the self-dual 
string of six-dimensional supergravity. Here we are interested only in 
one-dimensional extended object with horizons (black strings). The black strings
are realized only   for three, five and six dimensions \cite{Gibb}. 
In \cite{Lee}, we have already investigated the propagation of string fields in the 
three-dimensional extremal black string background. Unlike the higher
 dimensional cases, the exact conformal field theory is known
 and thus the particle contents can be determined.

\section{Five-dimensional black string}

We start with  the  five-dimensional Einstein-Maxwell action, 

\begin{equation}
S_{5d} = \int d^5 x \sqrt{-g} 
  ( R - {1 \over 4} F^2_{MN} ),
\end{equation}
where $x^M$ are the coordinates$(t,x,r,\theta,\phi)$ for five dimensions.
The equations of motion are given by
\begin{equation}
R_{MN} + {1 \over 12} F^2 g_{MN} - {1 \over 2} F_{MP} F^P_{~N} = 0,
\end{equation}

\begin{equation}
\nabla_M F^{MN} = 0.
\end{equation}
The static black string solution to the above equations in the 
extremal limit  is given by 
\begin{equation}
 \bar g_{MN} =
 \left(  \begin{array}{ccccc} - (1 - {\mu \over r}) & 0 & 0 & 0 & 0  \\
             0 & (1 - {\mu \over r}) & 0 & 0 & 0  \\
             0 & 0 &  (1 - {\mu \over r})^{-2}  & 0 & 0  \\
             0 & 0 & 0 & r^2 & 0  \\
             0 & 0 & 0 & 0 & r^2 \sin^2 \theta^2  
\end{array}   \right),
\end{equation}
\begin{equation}
\bar F_{\theta\phi} = Q \sin \theta.
\end{equation} 
Here the constant $\mu$  is related the  ADM mass per unit length of 
the string ($M$) 
and magnetic charge per unit length ($Q$)  as
\begin{equation}
\mu = {2|Q| \over \sqrt{3}} = {4 M \over 3}.
\end{equation}
Eq.(15) represents a straight, static black string which is an one-dimensional
extended object with horizon at $r=\mu$.
The above metric is not only translationally invariant, but also boost invariant.
This allows us to introduce two null-coordinates ($v= x +t, u = x -t$) and 
null Killing vector field ($ { \partial \over \partial v}$). Using these, 
the line element can be rewritten as
\begin{equation}
ds^2_{4+1}= ( 1 -{ \mu \over r}) du dv  + 
( 1 -{  \mu \over r})^{-2}dr^2 + r^2 d \Omega^2_2,
\end{equation}
where $d\Omega^2_2$ denotes the metric of the unit two-sphere.
After the double dimensional reduction of this balck string,
one can find  a four-dimensional
extremal dilaton black hole with $a = 1/ \sqrt 3$,
\begin{equation}
ds^2_{4}= -( 1 -{ \mu \over r})^{{ 3 \over 2}} dt^2  + 
( 1 -{  \mu \over r})^{-{ 3 \over 2}}dr^2 + 
( 1 -{  \mu \over r})^{{ 1 \over 2}}r^2 d \Omega^2_2,
\end{equation}

For our purpose, we introduce the  perturbation fields $({\cal F}(x), h(x))$ 
around the black string background  as\cite{Garf2,Lee}
\begin{eqnarray}
&&F_{\theta \phi} = \bar F_{\theta \phi} + {\cal F}_{\theta \phi} 
                  = \bar F_{\theta \phi} ( 1 + {\cal F} ),    \\
&&g_{MN} = \bar g_{MN} + h_{MN}= 
\bar g_{MN} + h k_M k_N ,
\end{eqnarray}
where $k_M$ is the null, orthogonal 
Killing vector which satisfies
\begin{equation}
k^M k_M =0, ~~\bar \nabla_{(M}k_{N)} =0, 
~~ k_{[M}\bar \nabla_{N} k_{L ]} =0 .
\end{equation}
Here we choose $k^M = ({\partial \over \partial v})^M$.
For this metric perturbation, 
the harmonic gauge condition 
($\bar \nabla_M h^M_{~N} = {1 \over 2} \bar \nabla_N h^M_{~M}$) is trivially 
satisfied.
This ansatz for the metric perturbation is valid for the extremal black strings, 
but not for the extremal black holes. This is because only the extremal balck string
has the null Killing symmetry.

In order to obtain the  equations governing the perturbations, one has to 
linearize (13) and (14) as 

\begin{equation}
\delta R_{MN} (h) + {1 \over 6} \bar F_{\theta\phi} \bar F^{\theta\phi}
   (h_{MN} + 2 \bar g_{MN} {\cal F})
- ( \bar F_{M \theta} \bar F_N^{~\theta} + \bar F_{M \phi}\bar F_N^{~\phi} )
 {\cal F}= 0,
\end{equation}
\begin{equation}
\bar \nabla_M ( \bar F^{MN} {\cal F} ) +
\delta \Gamma^M_{MP} (h) \bar F^{PN} =0, 
\end{equation}
where  
\begin{equation}
\delta R_{MN} (h)  = - {1 \over 2} 
(\bar \nabla_M \bar \nabla_N h^P_{~~P}
 +\bar \nabla^P \bar \nabla_P h_{MN}   
 -  \bar \nabla^P \bar \nabla_N h_{MP}   
 -  \bar \nabla^P \bar \nabla_M h_{NP}),
\end{equation}
\begin{equation}
\delta \Gamma^M_{NP} (h)  = {1 \over 2} \bar g^{ML} 
( \bar \nabla_P h_{NL} + \bar \nabla_N h_{PL} - 
\bar \nabla_L h_{NP} ).
\end{equation}
From (24), taking $N= \theta$ and $\phi$ respectively, one can easily find 
\begin{equation}
\partial_\theta {\cal F} = 0, ~~~~~~~~ \partial_\phi {\cal F} = 0.
\end{equation}
This implies that if ${\cal F} \not=0$, ${\cal F}$ depends only on
$r,~t$ and $x$ as  
\begin{equation}
{\cal F} = {\cal F}(r,t,x).
\end{equation}

From (23), one finds fifteen equations. However, only nine of them are nontrivial 
and these  are given by 
\begin{eqnarray}
(v,u) : & \partial_v^2 h + { 2 \mu^2 \over  r^4} {\cal F} = 0, \\
(u,u) : & (r - \mu)^2 \partial^2_r h 
         +   {2(r^2 - \mu^2) \over r }  \partial_r h +  {2\mu^2 \over r^2} h
         +  \partial^2_\theta h + \cot \theta \partial_\theta h 
         + { 1 \over \sin^2 \theta} \partial^2_\phi h = 0, \\
(u,r) : & ( \partial_r  - { \mu \over r (r-\mu)} ) \partial_v (k_u^2 h) = 0, \\
(u,\theta) : & \partial_\theta \partial_v h = 0,  \\ 
(u,\phi) : & \partial_\phi \partial_v h = 0,  \\ 
(r,r) : &  {\cal F} = 0, \\
(\theta,\theta) : &  {\cal F} = 0, \\
(\phi,\phi) : &  {\cal F} = 0,
\end{eqnarray}
From (27) and (34)-(36), we obtain ${\cal F}=0$. This means that the Maxwell 
mode is not the propagating one in the five-dimensional black string background.
Substituting this into (29) and (31)-(33) leads to
\begin{equation}
 \partial_v h=0.
\end{equation}
In order to solve one remaining (30), let us define $h$ as 
$h \equiv p(r) h'(r) U(u) Y( \theta, \phi)$.
Here the spherical harmonics $Y( \theta, \phi)$ on $S^2$ satisfies
the angular eigenvalue equation
\begin{equation}
{\partial^2 Y \over \partial \theta^2}
+ \cot \theta {\partial Y \over \partial \theta}
+ {1 \over \sin^2 \theta} {\partial^2 Y \over \partial \phi^2}
= - l (l +1) Y,
\end{equation}
where $l$ is a non-negative integer.
Then for  $p(r)= r /(r-\mu)$, (30) is reduced to
\begin{equation}
 (r - \mu)^2 \partial^2_r h' + 2 (r - \mu) \partial_r h' - l(l+1) h' = 0.
\end{equation}
Assuming the form of solution $h'=(r- \mu)^{\alpha}$,
the general solution to this is
\begin{equation}
 h' = C_1 (r - \mu)^{l} + {C_2 \over (r - \mu)^{l +1}}.
\end{equation}
At this stage we can compare our results with
that of three-dimensional black string.
The $l=0$ case cooresponds to the three-dimensional black string. 
In this case, from (39) one finds
\begin{equation}
 \partial^2_r h' + {2 \over (r - \mu)} \partial_r h' = 0.
\end{equation}
This is exactly the equation for three-dimensional black string and
the corresponding solution is given by $h'_{3d}= { C_2 \over r - \mu}$\cite{Garf2,Lee}.
Therefore we can interpret the five-dimensional black string as the direct product
of three-dimensional black string and two-sphere ($S^2$).

\section{Six-dimensional  black string}
 
The six-dimensional  action from (1) is given by

\begin{equation}
S_{6d} = \int d^6 x \sqrt{-g} 
  ( R - {1 \over 3} H^2_{MNP} ),
\end{equation}
where $x^M$ are the coordinates$(t,x,r,\chi,\theta,\phi)$ for six dimensions
and an $H_{MNP}$ is the three-form $H=dB$.
The equations of motion are given by
\begin{equation}
R_{MN} + {1 \over 6} H^2 g_{MN} -  H_{MPQ} H_M~^{PQ} = 0,
\end{equation}

\begin{equation}
\nabla_M H^{MPQ} = 0.
\end{equation}

The static black string solution to the above equations in the 
extremal limit  is given by 
\begin{equation}
ds^2_{5+1} = -\Big[ 1 -({ \mu \over r})^2 \Big] dt^2  + \Big[ 1 -({ \mu \over r})^2 \Big] dx^2
+ \Big[ 1 -({  \mu \over r})^2 \Big]^{-2}dr^2 + r^2 d \Omega^2_3,
\end{equation}
\begin{equation}
\bar H_{\chi\theta\phi} = Q \sin^2 \chi \sin \theta,
\end{equation} 
where $d \Omega^2_3$ represents the metric for  three sphere $(S^3)$.
Here the constant $\mu$  is related to the magnetic charge per unit length ($Q$)  in the extremal limit as
\begin{equation}
\mu^4 = { Q^2 \over 2}.
\end{equation}
Eq.(45) represents a six-dimensional, static black string. 
By the same argument as in the five-dimensional case, one can 
introduce two null-coordinates ($v= x +t, u = x -t$) and 
null Killing vector field ($ { \partial \over \partial v}$).  
The line element can be rewritten as

\begin{equation}
ds^2= \Big[ 1 -({ \mu \over r})^2 \Big] du dv  
+ \Big[ 1 -({  \mu \over r})^2 \Big]^{-2}dr^2 + r^2 d \Omega^2_3.
\end{equation}
After the double dimensional reduction of this balck string,
one can find  a five-dimensional
extremal dilaton black hole with $a = \sqrt {2/3}$,
\begin{equation}
ds^2_{5}= \Big [ 1 -({ \mu \over r})^2 \Big ]^{ 1 \over3 }
\Big [-( 1 -({ \mu \over r})^2) dt^2  
+ ( 1 -({  \mu \over r})^2)^{-2}dr^2 + r^2 d \Omega^2_3 \Big ].
\end{equation}
Except the conformal factor,
this metric is  similar to the five-dimensional Reissner-Nordstr\"om black
hole, which was considered in studying microscopic origin of the Bekenstein-
Hawking entropy\cite{Vafa}. 

Also one can introduce the  perturbation fields $({\cal H}, h)$ 
around the extremal black string background  as\cite{Lee,myung}
\begin{eqnarray}
&&H_{\chi \theta \phi} = \bar H_{\chi \theta \phi} + {\cal H}_{\chi \theta \phi} 
                  = \bar H_{\chi \theta \phi} ( 1 + {\cal H} ),    \\
&&g_{MN} \equiv \bar g_{MN} + h_{MN}= 
\bar g_{MN} + h k_M k_N ,
\end{eqnarray}
where $\bar g_{MN}$ represents (45) and $k_M$ is the null, orthogonal 
Killing vector in six dimensions.
After linearizing (43) and (44), we have 

\begin{equation}
\delta R_{MN} (h) + {1 \over 6} \bar H^2(h_{MN} + 2 \bar g_{MN} {\cal H})
-2 \bar H_{MPQ} \bar H_N~^{PQ}{\cal H}= 0,
\end{equation}
\begin{equation}
\bar \nabla_M ( \bar H^{MNP} {\cal H} ) +
\delta \Gamma^M_{MQ} (h) \bar H^{QNP} =0. 
\end{equation}
From (53), taking $N= \chi$, $\theta$ and $\phi$ respectively
 one finds 
\begin{equation}
\partial_\chi {\cal H} = 0, ~~~~~~~~ \partial_\theta {\cal H} = 0,
~~~~~~~~~\partial_\phi {\cal H} = 0.
\end{equation}
This implies that if ${\cal H} \not=0$, ${\cal H}$ depends only on
$r,~t$ and $x$,  
\begin{equation}
{\cal H} = {\cal H}(r,t,x).
\end{equation}

From (52), one finds twenty-one equations. However,  ten nontrivial 
equations  are given by 
\begin{eqnarray}
(v,u) : & \partial_v^2 h + { 8 \mu^4 \over  r^6} {\cal H} = 0, \\
(u,u) : &{ (r^2 - \mu^2)^2 \over r^4 } \partial^2_r h 
         +   {(r^2 - \mu^2) (3 r^3 + 5 \mu^2) \over r^5 } \partial_r h 
         +  {8 \mu^4 \over r^6} h               \nonumber      \\
        & + { 1 \over r^2 } ( \partial^2_\chi  
              + {\partial^2_\theta \over  \sin^2 \chi}  
              + {\partial^2_\phi \over  \sin^2 \chi \sin^2 \theta}  
              + 2 \cot \chi \partial_\chi  
              + { \cot \theta \over \sin^2 \chi } \partial_\theta ) h = 0, \\ 
(u,r) : & ( \partial_r  - { 2 \mu \over r (r^2-\mu^2)} ) 
                      \partial_v (k_u^2 h) = 0, \\
(u,\chi) : & \partial_\chi \partial_v h = 0,  \\ 
(u,\theta) : & \partial_\theta \partial_v h = 0,  \\ 
(u,\phi) : & \partial_\phi \partial_v h = 0,  \\ 
(r,r) : &  {\cal H} = 0, \\
(\chi,\chi) : &  {\cal H} = 0, \\
(\theta,\theta) : &  {\cal H} = 0, \\
(\phi,\phi) : &  {\cal H} = 0.
\end{eqnarray}
From (54) and (62)-(65), we obtain ${\cal H}=0$. This means that the three-form
${\cal H}_{\chi\theta\phi}$ is not the propagating one in 
the six-dimensional black string background.
Putting this into (56) and (58)-(61), we have 
\begin{equation}
 \partial_v h=0.
\end{equation}
In order to solve (57), we define $h$ as 
$h \equiv q(r) h''(r) U'(u) Y(\chi, \theta, \phi)$,
where the spherical harmonics $Y(\chi, \theta, \phi)$ on $S^3$ satisfies
the angular eigenvalue equation
\begin{equation}
{\partial^2 Y \over \partial \chi^2} 
+ 2\cot \chi {\partial Y \over \partial \chi} 
+ { 1 \over \sin \chi^2}{\partial^2 Y \over \partial \theta^2}
+ { \cot \theta \over \sin \chi^2}{\partial Y \over \partial \theta}
+ { 1 \over \sin^2 \chi \sin^2 \theta} {\partial^2 Y \over \partial \phi^2}
= - L (L +2) Y,
\end{equation}
where $L(L+2)$ a  eigenvalue for the Laplacian operator on the three
sphere.
Choosing  $q(r)={ r^2 \over r^2-\mu^2}$, then (57) is reduced to
\begin{equation}
 {(r^2 - \mu^2)^2 \over r^4} \partial^2_r h'' + 
{ (r^2 - \mu^2)(3r^2 + \mu^2) \over r^5} \partial_r h''
 - { L(L+2) \over r^2} h'' = 0.
\end{equation}

In order to solve the above equation, we introduce a new  variable $(z)$
as
\begin{equation}
z \equiv { r^2 \over \mu^2} -1.
\end{equation}
Then (68) can be rewritten as
\begin{equation}
4 \partial^2_z h''(z) + 
{ 8 \over z} \partial_z h''(z)
 - { L(L+2) \over z^2} h''(z) = 0.
\end{equation}
Assuming the form of solution $h''(z)=z^{\alpha}$, one finds
\begin{equation}
\alpha= { L \over 2},~~~~~~ -{ L+2 \over 2}.
\end{equation}
The general solution to (68) is
\begin{equation}
 h''(r) = C'_1 ({r^2 - \mu^2 \over \mu^2})^{L} +
C'_2 ({r^2 - \mu^2 \over \mu^2})^{-{L+2 \over 2}}
\end{equation}
with two arbitray constants $C'_1$ and $C'_2$.

\section{Discussion}

Let us first discuss  the stability of higher-dimensional black strings.
Conventionally, in deciding whether a black hole is stable or not, we 
start with a perturbation which is regular everywhere in space at the initial time
$t$. In the cases of most black holes, the linearized equation which
governs the perturbation is the Schr\"odinger-type equation\cite{Chan,Kim}.
 And we then investigate whether such a perturbation grows with time.
If there exists an exponentially growing mode, the black hole is unstable.
This implies that one finds a physical mode with the potential well 
around the black hole. The potential barrier implies the scattering state, while
the potential well gives us the scattering state as well as the bound state.
It is well known that the bound state solution takes the form of  exponentially
increasing or decreasing functions.

Here we have  no additional 
constraint for determining $U(x,t)$ except the chiral constraint: $\partial_v U(x,t)=0$ within this
scheme.
We assume the normal mode solution of the form
\begin{equation}
U(x,t)= e^{-iEt} e^{-i\Pi x}.
\end{equation}
From $\partial_v U(x,t)=0$, one finds $\Pi= -E$. Then the form 
of $U(x,t)$ is determined as
\begin{equation}
U(u)= e^{i \Pi u}.
\end{equation}
This is a plane wave along the $v$=constant null line.
And this is called the longitudinal wave, since it carries only momentum $\Pi$ 
along the string direction ($x$-direction) \cite{Horo4}. 
 Hence the graviton mode ($ h=p(r)h'(r)U(u)Y_{lm}(\theta, \phi)$)can be the 
propagating wave in the black string background.

On the other hand, we may choose $\Pi= i\alpha$. Then (74) leads to an exponentially
growing mode $\tilde U(u)= e^{ \alpha (t-x)}$ with respect to the time.
Applying the argument of black hole stability to the extremal black strings, one finds that
it is unstable. This is because there is no restriction on choosing
$\Pi$ as either real or imaginary within the extremal black strings.
This point  contradicts  to that of the black holes. Namely, one  obtained 
the Schr\"odinger-type equation for the $a={ 1 \over \sqrt 3}$ extremal
dilaton balck hole\cite{Hol}, whereas we cannot obtain
the corresponding equation for the five-dimensional extremal black string.
 This is a result of our setting
of $h_{MN}$ in (21). Although one can easily obtain the solutions from this choosing
of $h_{MN}$, one cannot find the Schr\"odinger-type equation for the extremal black strings.

Next, we consider the problem of classical hair in the black string theory.
The classical no-hair theorem of general relativity severely restricts the 
kinds of fields that  can exist outside a black hole in four-dimensional
spacetime. For example, in a static geometry with a smooth event horizon
the only field which is well behaved both in the asymptotically flat region
 and on the event 
horizon is monopole gravitational and electromagnetic fields.
Here we define a black string with hair to be   a geometry with a smooth
event horizon and a  field which is nonsingular at that horizon.
Conventionally, one takes a static
solution to the linearized equations as a kind of hair\cite{Lars}.
In this case we require that the static solution
 be smooth both at the horizon $(r=\mu)$ and at spatial infinity
$(r= \infty)$\cite{Cole}. Let us consider our case. We note that (39) and (68),
as they stand, are static equations for the graviton.
 Then the full static solutions for graviton are given by
\begin{equation}
 p(r)h'(r) = C_1 r(r - \mu)^{l-1} + {C_2 r \over (r - \mu)^{l +2}},
\end{equation}
\begin{equation}
 q(r)h''(r) =C'_1 ({r^2 - \mu^2 \over \mu^2})^{L-1} +
C'_2 ({r^2 - \mu^2 \over \mu^2})^{-{L+4 \over 2}}.
\end{equation}

The first terms in (75) and (76)  are smooth near the horizon and diverge as $r \to \infty$,
while the second terms are singular on the horizon and converge in the asymptotically
flat region. Thus one cannot find a smooth solution which behaves well on
both limits. This means that in the strict sense of black hole hair,
there is no hair in the five and six-dimensional extremal black strings. 

In conclusion, we found the solutions which propagate in the 
higher dimensional extremal black string backgrounds. These correspond to the 
graviton mode. Both the Maxwell$({\cal F})$ and three-form $({\cal H})$ fields
are non-propagating modes.
This seems to be a controversial result, compared with the black holes. 
We consider the conventional counting of degrees of freedom.
The number of degrees of freedom for the gravitational field ($h_{\mu\nu}$) in 
$D$-dimensions is $(1/2) D (D -3)$.
 We have $-1$ for $D=2$. This means that in two dimensions
the contribution of graviton is equal and opposite to that of a spinless particle (dilaton).
In the 2d dilaton black hole, two graviton-dilaton modes are thus trivial gauge
artefacts \cite{Kim}.
In the case of $D =3$, we have no  propagating gravitons\cite{myung}.  For $D=4$ Schwarzschild black
hole, we obtain two degrees of freedom. These correspond to Regge-Wheeler mode
 for odd-parity perturbation and Zerilli mode for even-parity perturbation \cite
{Chan,Kwon}. In the cases of $D=5,6$, one finds five and nine degrees of freedom respectively. 
However, it is emphasized that this is a conventional counting which is
suitable for the black holes. We note that our model is the extremal
black string with the null Killing symmetry. In this case we have only one propagating graviton
in five dimensions  and also one graviton in six dimensions.
Furthermore, the Maxwell  and three-form field have $D-2$ and $(D-2)(D-3)/2$ degrees of freedom
 respectively.
A naive counting gives us  3 for the five-dimensional Maxwell field
and 6 for the six-dimensional three-form field.  However, it turns out that
both the Maxwell $({\cal F})$ and three-form $({\cal H})$ fields
are non-propagating modes in the extremal black string backgrounds.
It seems that the conventional counting for degrees of freedom is not suitable for the extremal
black strings.

\acknowledgments

This work was supported in part by the Basic Science Research Institute 
Program, Ministry of Education, Project NOs. BSRI--96--2441, BSRI--96--2413
and by Inje Research and Scholarship Foundation.


\newpage

\begin{references}

\bibitem{Vafa} A. Strominger and C. Vafa, Phys. Lett. {\bf B379}, 99 (1996).
\bibitem{Callan} C. Callan and J. Maldacena, Nucl. Phys. {\bf B472}, 591 (1996).
\bibitem{Horo1} G. Horowitz and A. Strominger, hep-th/9602051.
\bibitem{Breck} J. Breckenridge, R. Myers, A. Peet and C. Vafa, hep-th/9602065.
\bibitem{John} C. Johnson, R. Khuri, and R. Myers, hep-th/9603061.
\bibitem{Horo2} G. Horowitz, D. Lowe, and J Maldacena,  Phys. Rev. Lett. {\bf 77}, 430 (1996).
\bibitem{Horo3} G. Horowitz, gr-qc/9604051.
\bibitem{Gibb} G.W. Gibbons, G.T. Horowitz, Class. Quant. Grav. {\bf 12}, 297 (1995).
\bibitem{green}  M. Green, J. Schwarz, and E. Witten, {\it Superstring theory I,II}
                (Cambridge Univ. Press, New York, 1987).
\bibitem{Horo4} G. Horowitz  and D. Marolf, hep-th/9605224, hep-th/9606113.
\bibitem{Polch} J. Polchinski,  Phys. Rev. Lett. {\bf 75}, 4724 (1995).
\bibitem{Garf1} D. Garfinkle and T. Vachaspati, Phys. Rev. {\bf D42}, 1960
                 (1990).                
\bibitem{Garf2} D. Garfinkle, Phys. Rev. {\bf D46}, 4286 (1992).    
\bibitem{Tsey1} G. Horowitz and A. Tseytlin,  Phys. Rev. {\bf D50}, 5204 (1994).
\bibitem{Tsey2} G. Horowitz and A. Tseytlin,  Phys. Rev. {\bf D51}, 2896 (1995).        
\bibitem{Chan} S. Chandrasekhar, {\it The Mathematical Theory of Black Hole}
                (Oxford Univ. Press, New York, 1983). 
\bibitem{Kim}  J.Y. Kim, H. W. Lee and Y. S. Myung, Phys. Lett. 
               {\bf B328}, 291 (1994) ;Y. S. Myung, Phys. Lett. {\bf B334}, 29 (1994);
               {\it ibid.} {\bf B362}, 46 (1995).
\bibitem{Lee}   H. W. Lee, Y. S. Myung, J. Y. Kim, and D. K. Park, hep-th/9607001.             
\bibitem{myung} H. W. Lee, Y. S. Myung, and J. Y. Kim, Phys. Rev. {\bf D52}, 2214 (1995).
\bibitem{Hol}  C. Holzhey and F. Wilczek, Nucl. Phys. {\bf B380}, 447 (1992).
\bibitem{Lars}  F. Larsen and F. Wilczek, hep-th/9604134; hept-th/9609084.
\bibitem{Cole} C. Coleman, J. Preskill, and F. Wilczek, Nucl. Phys.{\bf B378}, 175 (1992);
               A. Peet, L. Susskind, and L. Thorlacius, Phys. Rev. {\bf D48}, 2415 (1993).
\bibitem{Kwon} O. J. Kwon, Y. D. Kim, Y. S. Myung, B. H. Cho and Y. J. Park,
               Phys. Rev. {\bf D34}, 333 (1986); Int. J. Mod. Phys. {\bf A1}, 701 (1986). 
\end{references}
\end{document}